\documentclass[envcountreset,envcountsect,envcountsame]{llncs}
\usepackage{latexsym}

\begin{document}

\newcommand{\I}[1]{\mbox{{\it #1}}}
\newcommand{\T}[1]{\fbox{\rule[-0.5ex]{0ex}{2ex}\tt #1}}
\newcommand{\BQ}{\begin{quote}}
\newcommand{\EQ}{\end{quote}}
\newcommand{\BI}{\begin{itemize}}
\newcommand{\EI}{\end{itemize}}
\newcommand{\BE}{\begin{enumerate}}
\newcommand{\EE}{\end{enumerate}}
\newcommand{\BD}{\begin{definition}}
\newcommand{\ED}{\end{definition}}
\newcommand{\BL}{\begin{lemma}}
\newcommand{\EL}{\end{lemma}}
\newcommand{\BEX}{\begin{example}}
\newcommand{\EEX}{\end{example}}

\newcommand{\BV}{\begin{verbatim}}
\newcommand{\BC}{\begin{center}}
\newcommand{\EC}{\end{center}}

\newcommand{\calc}{$\lambda_{ndlr}~$}
 
\newcommand{\IT}{\item}
\newcommand{\bottom}{\perp}
\newcommand{\derive}{\vdash}
\newcommand{\impl}{\Rightarrow}
\newcommand{\SB}[1]{{[\![#1]\!]}}
\newcommand{\SBpsi}[1]{{\SB{#1}_\psi}}

\newcommand{\append}{+\hspace{-4pt}+}

\newcommand{\CON}{{\Downarrow}}
\newcommand{\NOTCON}{{\not\Downarrow}}
\newcommand{\CONA}{{\Downarrow\!\Downarrow}}
\newcommand{\NOTCONA}{{-\!\!\!-\!\!\!-\!\!\!\!\!\!\!\!\!\!\!\Downarrow\!\Downarrow~Ê}}

\newcommand{\RA}{\longrightarrow}
\newcommand{\LA}{\longleftarrow}

\newcommand{\LRA}{\Longrightarrow}

\newcommand{\RAone}{\stackrel{1}{\rightarrow}}
\newcommand{\RAST}{\stackrel{*}{\rightarrow}}

\newcommand{\RAN}{\stackrel{n}{\rightarrow}}
\newcommand{\letset}{\stackrel{set}{\longrightarrow}}
\newcommand{\letsetst}{\stackrel{set,*}{\longrightarrow}}

\newcommand{\RAP}[1]{\stackrel {#1}{\RA}}
\newcommand{\LAP}[1]{\stackrel {#1}{\LA}}

\newcommand{\RRAP}[1]{\stackrel {#1}{\!-\!\!\!\RA}}
\newcommand{\RRRAP}[1]{\stackrel {#1}{\!-\!\!\!-\!\!\!\RA}}
\newcommand{\LLAP}[1]{\stackrel {#1}{\LA \!\!\!-\!}}
\newcommand{\LLLAP}[1]{\stackrel {#1}{\LA \!\!\!-\!\!\!-\!\!\!-\!}}
\newcommand{\LLLLAP}[1]{\stackrel {#1}{\LA \!\!\!-\!\!\!-\!\!\!-\!\!\!-\!\!\!-\!}}
\newcommand{\LLLLLAP}[1]{\stackrel {#1}{\LA \!\!\!-\!\!\!-\!\!\!-\!\!\!-\!\!\!-
\!\!\!-\!\!\!-\!\!\!-\!\!\!-\!\!\!-\!}}

\newcommand{\RRA}{\Rightarrow}
\newcommand{\BT}{\begin{tabular}{p{1cm}cp{5.5cm}l}}
\newcommand{\ET}{\end{tabular}}

\newcommand{\ab}{\hspace*{1cm}}
\newcommand{\ah}{\hspace*{.5cm}}

\newcommand{\botelem}{\bot\!\!\!\in}
\newcommand{\topelem}{\top\!\!\!\!\in}

\newcommand{\pprime}{{\prime\prime}}
\newcommand{\ppprime}{{\prime\prime\prime}}
\newcommand{\tbot}{{\tt Bot}}
\newcommand{\ttop}{{\tt Top}}
\newcommand{\tnil}{{\tt Nil}}
\newcommand{\ttlc}{{\tt lcase}}
\newcommand{\ttrue}{{\tt True}}
\newcommand{\tfalse}{{\tt False}}
\newcommand{\tlet}{{\tt let}}
\newcommand{\tletrec}{{\tt letrec}}
\newcommand{\tchoi}{{\tt choice}}
\newcommand{\tcase}{{\tt case}}
\newcommand{\tletx}[3]{(\tlet~#1 = #2~{\tt in}~#3)}
\newcommand{\tletrx}[2]{(\tletrec~#1 ~{\tt in}~#2)}
\newcommand{\LAM}[2]{{(\lambda~#1~.~#2)}}
\newcommand{\QED}{{$\Box$}}

\newcommand{\hspv}{\hspace*{4mm}}
\newcommand{\hspvv}{\hspace*{8mm}}

\newcommand{\comment}[1]{}
\bibliographystyle{alpha}
\sloppy
\parindent0em



\author{Manfred Schmidt-Schau{\ss} and Michael Huber}
\institute{Fachbereich Informatik \\
Johann Wolfgang Goethe-Universit{\"at}\\
Postfach 11 19 32\\
D-60054 Frankfurt, Germany\\
E-mail: {\tt schauss@ki.informatik.uni-frankfurt.de}}


\title{A Lambda-Calculus with  letrec,  case, constructors and non-determinism}
\maketitle

\begin{abstract}
A non-deterministic call-by-need lambda-calculus \calc with case, constructors,  
letrec and a (non-deterministic) erratic choice, based on 
rewriting rules is investigated. A standard reduction is defined as a variant of left-most outermost reduction.
The semantics is defined by contextual equivalence  of expressions instead of using 
$\alpha\beta(\eta)$-equivalence.     
It is shown that several  program transformations are correct,
for example all (deterministic) rules of the calculus, and in addition the rules 
for garbage collection, removing indirections 
and unique copy. 

This shows that the combination of a context lemma and a meta-rewriting on reductions 
using complete sets of commuting (forking, resp.) diagrams is a useful and successful method for 
providing a semantics of a functional programming language and proving correctness  
of program transformations.  
\end{abstract}

\section{Introduction}

Functional programming languages are based on extended lambda calculi and the corresponding
rewrite semantics.  There are several methods of giving these languages a semantics and proving the correctness
of program transformations: 
\BI
\item A {\em denotational semantics} uses a mathematical domain and a mapping from expressions to their 
denotation. This defines an equivalence of expressions, which can be used to define a notion of correctness
of program transformations. This area is well-developed, but reaches its limits if  non-deterministic 
operations are possible in the language. 
\item An {\em operational semantics} defining the evaluation of expressions (the execution, resp.).  Sometimes this 
is used with a kind of syntactic equality (e.g. $\alpha\beta(\eta)$-equality in the 
lambda-calculus). It 
could also be complemented by a behavioral equivalence, which can be used  to define the a  notion of correctness of 
program transformation. 
\item A {\em contextual semantics} is a kind of operational semantics as above enhanced with an 
approximation relation based on a contextual preordering (see e.g. \cite{smith:92,mason-smith-talcott:96,pitts:97}).
An expression $s$ has less information than an expression $t$, iff in all contexts $C[]$, if $C[s]$ gives some information
(e.g. terminates),
then $C[t]$ also gives some information (i.e. terminates).  
This notion is directly adapted to define a notion of correctness of program transformation.
Often it  gives the  intuitive correct notion of program equivalence, and hence also of correct 
program transformations.
 \EI

The advantage of the contextual semantics is that the number of equality relations is maximal and that 
the derived properties are independent of a specific domain. The properties of the contextual preorder are comparable
to the orderings in domains; for example it is possible to  use fixed-point constructions for recursion. 
The contextual semantics is superior to the more syntax-oriented $\alpha\beta(\eta)$-equivalence,
since contextual semantics permits considerably more program transformations. 
    
An advantage of contextual semantics over the denotational approach becomes obvious  if non-determinism is 
on board and also sharing in the form 
of a (non-recursive or recursive) {\tt let}. It appears to be very hard to construct a useful domain for 
denotational semantics
in the presence of non-determinism and higher-order functions, 
whereas it is possible to 
use the contextual equivalence for defining an intuitive correct semantics. This 
can then be used to prove correctness of program transformations
sometimes exploiting rewriting techniques. 
A slight disadvantage  of the contextual semantics (w.r.t. economy of proofs) is that it depends on 
the available syntactic constructs, hence on the set of contexts, and the 
defined standard reduction.

The prominent syntactic property of the lambda-calculus is confluence of reduction \cite{barendregt:84}.
In the framework of a contextual semantics for the lambda-calculus (see e.g. \cite{abramsky-lazy:90}),
confluence is not thus important and is replaced by the correctness of program transformations. 
The really interesting propositions are:
\BI
\item Every beta-reduction transforms a program $P$ into an equivalent one $P'$, 
  meaning that $P$ and $P'$ are contextually equivalent. This is the required modification of confluence.  
\item (standardization) Whenever there is a reduction of an expression $t$ to an abstraction, then the standard reduction 
terminates, i.e. reduces $t$ to an abstraction.
\EI
These properties can be generalized to extended lambda-calculi, where confluence may be false 
(see e.g. \cite{ariola-klop:94}), but  contextual equivalence can be easily adapted.     

Another approach is Rewriting Logic (see e.g. \cite{Meseguer:00}), which is a step in the direction of 
providing a semantics for
programming languages based on rewriting rules. This appears to work for deterministic languages based
on rewriting rules. However, the contextual semantics is our method of choice for the non-deterministic case.
\vspace{0,3cm} 

In this paper we present the calculus \calc that is rather close to a non-strict functional core language. 
Reduction is 
like lazy call-by-need evaluation in functional programming languages. \calc can be seen as a 
generalization of the calculus in
\cite{kutzner:98,kutzner:00} and thus of the calculi in \cite{ariola:95,ariola:97,maraistoderskywadler:98}, 
which treat sharing in the lambda calculus. It also is a generalisation of \cite{moran-sands-carlsson:99} 
insofar as the 
language of expressions is not restricted to have only variables as arguments in applications. The calculus \calc 
is
related to the calculus in \cite{fundio:00}, where a similar language is investigated, but with the emphasis on
an IO-interface.  
 
Another method for treating sharing are explicit substitutions \cite{abadi:91}, which optimize resource usage 
of reductions by exploiting sharing, 
however, it is i) based on $\alpha\beta(\eta)$-equivalence and ii) the reduction rules 
 are in general not compatible with non-determinism, i.e. not with \calc nor with the calculus in \cite{kutzner:00}; 
in particular, the let-over-lambda-rules are incompatible with non-determinism. 

Specific ingredients of \calc are
\BI
\item sharing by using {\tt letrec}, which moreover allows recursive definitions.
\item a non-deterministic (erratic) {\tt choice}, which allows to choose between two expressions.
\item a modified beta-reduction that prevents an unwanted duplication of non-deterministic expressions.  
\EI

The motivation to investigate non-determinism is to model interfaces of lazy functional languages to the outside 
world, 
i.e. to model input/output. This is done by a simulation of an IO-action by a nested \tchoi-expression
that represents the set of possible input values of the IO.      
  
The paper proposes to investigate extended lambda-calculi by using operational methods and a
contextual semantics. The contextual semantics includes a measure for the number of non-deterministic
steps. 
As a method for proving program transformation to be correct we propose to use complete sets of reduction diagrams in 
combination with
an appropriate context lemma.   

The results are that for \calc a rather large set of basic program transformations
is proved to be correct. The paper also demonstrates the power of the method, since the reduction rules of 
\calc are numerous and complex. 

As a check-program for complete sets, a program ``Jonah``
 was implemented to automatically test the complete reduction diagrams using a generate-and-test scheme; 
 Jonah can also be used to compute proposals for complete sets. 

As a remaining open problem the paper can be seen as a recommendation to start an  
investigation into adapting  the Knuth-Bendix method to automatically computing the reduction diagrams. 
However, the reduction diagrams for (llet) for example show that it would be necessary to integrate
a kind of meta-description like the Kleene-$*$.           

In this paper we do not present all proofs, but give enough hints and evidence of how the claims can be
verified and that they are valid.

\section{The calculus \calc}

The syntax of the language is as follows:\\
There is a set of type-names. 
For every type there are constructors $c$ coming with an arity $ar(c)$. 
This partitions the set of 
all constructors into the constructors belonging to different types. 
For a type $A$, $|A|$ is defined
to be the number of constructors belonging to $A$. The constructors belonging to type
$A$ are indexed, and $c_{A,i}$ denotes the $i^{th}$ constructor of type $A$.  

\begin{eqnarray*}
E &::=& V~|~C~|~(\tchoi~s~t)~|~(\tcase_A ~E~ Alt_1 \ldots Alt_{|A|})~|~(E_1~E_2)\\
& & |~(\lambda V. E)~|~\tletrx{V_1 = E_1, \ldots V_n = E_n}{E}\\
Alt &::=& (Pat \to E)\\
Pat &::=& c~V_1 ~\ldots ~V_{ar(c)}\\  
\end{eqnarray*}

where $E,E_i$ are expressions, $A$ is a type, $V,V_i$ are variables, and $C$ is a constructor.
The variables in a pattern $Pat$ must be different, and also new ones. Moreover, 
in a $\tcase_A$-expression, 
there is exactly one alternative  with a pattern of the form  
$(c_{A,i}~y_1 \ldots y_n)$  
for every constructor $c_{A,i}$. 
The constants $\tcase_A$ and $\tchoi$ can 
only occur
in a special syntactic construction. Thus expressions where $\tchoi$ or $\tcase_A$
is applied to a wrong number of arguments are not allowed.

The structure  $\tletrec$ obeys the following conditions: The variables in the bindings are
all distinct. We also assume that the bindings in $\tletrec$ are commutative, i.e. 
can be commuted without syntactic
change. $\tletrec$  is  recursive: I.e. the scope of $x_i$ in $\tletrx{x_i = E_i}{E}$ 
is $E_i,E$.  This allows to define closed, open expressions and $\alpha$-renamings. 
For simplicity we use the disjoint variable convention. I.e., all bound variables
in expressions are assumed to be disjoint. The reduction rules are such that 
the bound variables in the result are also made distinct by $\alpha$-renaming.
We also use the convention to omit parenthesis in denoting expression: $(s_1 \ldots s_n)$
denotes $(\ldots (s_1~s_2) \ldots  s_n)$.

We say that an expression of the form $(c~t_1 \ldots t_n)$ is a 
{\em constructor application}, if $n \le ar(c)$. A  constructor application  
of the form $(c~x_1 \ldots x_n)$ is called a 
{\em pure constructor application}.   
An expression of the form $(c~t_1 \ldots t_{ar(c)})$ is called a 
{\em saturated constructor application}.

\BD
Let $R, R^-$, be context classes defined as follows:  
\begin{eqnarray*}
R^- &::= &[]~|~R^-~E~|~(\tcase_A~R^-~alts) \\ 
R &::= &R^-~|~\tletrx{x_1 = E_1,\ldots,x_n = E_n}{R^-}\\
 & & |~\tletrx{x_1 = R^-_1[\cdot],x_2 = R^-_2[x_1],\ldots, x_j = R_{j}^-[x_{j-1}], \ldots }{R^-[x_j]}\\ 
& & \quad \mbox{where }  R^-_j \mbox{ is a context of class } R^- \\ 
\end{eqnarray*}
$R$ is called a {\em reduction context} and $R^-$ is called a {\em weak reduction context}.
For a term $t$ with $t = R^-[t_0]$, we say $R^-$ is {\em maximal}, iff there is no larger weak reduction context 
with this property. For a term $t$ with $t = R[t_0]$, we say the reduction context $R$ is {\em maximal}, iff  it is either a maximal
weak reduction context, or of the form 
$\tletrx{x_1 = R^-_1[\cdot],x_2 = R^-_2[x_1],\ldots, 
x_j = R_{j}^-[x_{j-1}], \ldots }{R^-[x_j]}$, where $t = \tletrx{x_1 = t_1, \ldots }{R^-[x_j]}$, 
$R^-_1[\cdot]$ is maximal for $t_1$,
and the number $j$ is maximal. 
\ED

For example the maximal reduction context of $\tletrx{x_2 = \lambda x. x, x_1 = x_2~x_1}{x_1}$ 
is $\tletrx{x_2 = [], x_1 = x_2~x_1}{x_1}$, in contrast to the non-maximal reduction context  
$\tletrx{x_2 = \lambda x. x, x_1 = x_2~x_1}{[]}$. 

The (call-by-need) reduction rules defined in \ref{def-red-rules} follow the principle of minimizing copying at the cost of perhaps 
following several indirections. This holds for the copy rule (cpn) as well as (case). The technical reason
is that this principle assures well-behaved reduction diagrams.


\begin{figure}[h]\label{figure-reductions}
\[\begin{array}{|ll|}\hline
\mbox{(lbeta)}&((\lambda x. s)~t)   \to   \tletrx{x = t}{s}\\
\mbox{(cpn)} & \tletrx{x_1 = s_1, x_2 = x_1, \ldots, x_j = x_{j-1}, x_{j+1} = s_{j+1} \ldots }{C[x_j]}  \\
& \quad \to \tletrx{x_1 = s_1, x_2 = x_1, \ldots, x_j = x_{j-1}, x_{j+1} = s_{j+1} \ldots }{C[s_1]}  \\
& \mbox{where } s_1 
     \mbox{ is an abstraction}\\
     
\mbox{(cpn)} & \tletrx{x_1 = s_1, x_2 = x_1, \ldots, x_j = x_{j-1}, x_{j+1} = C[x_j], \ldots }{s}  \\
& \quad \to \tletrx{x_1 = s_1, x_2 = x_1, \ldots, x_j = x_{j-1}, x_{j+1} = C[s_1], \ldots }{s}  \\
& \mbox{where } s_1 
     \mbox{ is an abstraction}\\
 \mbox{(llet)}& \tletrx{x_1 = s_1,\ldots,x_n = s_n}{\tletrx{y_1 = t_1,\ldots,y_m = s_m}{r}} \\
    &\to  \tletrx{x_1 = s_1,\ldots,x_n = s_n, y_1 = t_1,\ldots,y_m = s_m}{r}\\
\mbox{(llet)}& \tletrx{x_1 = s_1,\ldots,x_i =  \\
& \quad \quad \tletrx{y_1 = t_1,\ldots,y_m = t_m}{s_i}, \ldots,x_n = s_n}{r}  \\
    &\quad \to  \tletrx{x_1 = s_1,\ldots,x_n = s_n, y_1 = t_1,\ldots,y_m = s_m}{r}\\
\mbox{(lapp)} & (\tletrx{x_i = t_i}{t}~s)  \to   \tletrx{x_i = t_i}{(t~s)} \\
\mbox{(lcase)}& (\tcase_A~\tletrx{E}{t}~alts)  \to 
         \tletrx{E}{\tcase_A~t ~alts} \\      
 \mbox{(case)}& 
(\tcase_A~(c_{A,i}~t_1 \ldots t_n)~ \ldots 
((c_{A,i}~y_1 \ldots y_n) \to t) \ldots) \\
  & \to \quad 
  \tletrx{y_1 = t_1 \ldots  y_n = t_n}{t}\\ 
 \quad & \\
 \mbox{(case)}& 
    \begin{array}{ll}
       \tletrec & x_1 = (c_{A,i}~t_1 \ldots t_{j_1}),  \\
              & x_2 = x_1~t_{j_1+1} \ldots t_{j_2}, \\
              & \ldots \\
              & x_{m} = x_{m-1}~t_{j_{m-1}+1} \ldots t_{j_{m}}, \\
              & \ldots \\
              & C[\tcase_A~(x_{m}~t_{j_{m}+1} \ldots t_{j_{m+1}})~ \ldots ((c_{A,i}~z_1 \ldots z_n) \to t)])
     \end{array} \\
  &  \longrightarrow \\
  & \begin{array}{ll}
     \tletrec & x_1 = (c_{A,i}~y_1 \ldots y_{j_1}), y_1 = t_1, \ldots y_{j_1} = t_{j_1},  \\
          & x_2 = x_1~y_{j_1+1} \ldots y_{j_2}, y_{j_1+1} = t_{j_1+1}, \ldots, y_{j_2} = t_{j_2},\\
          & \ldots \\
          & x_m = x_{m-1}~y_{j_{m-1}+1} \ldots y_{j_m}, y_{j_{m-1}+1} = t_{j_{m-1}+1}, \ldots, y_{j_m} =t_{j_m}, \\
          & \ldots \\
          & C[\tletrx{y_{j_{m}+1} = t_{j_{m}+1},\ldots, y_{j_{m+1}} = t_{j_{m+1}}, z_1 = y_1, \ldots, z_n = y_n}{t}]
  \end{array} \\
   & \mbox{where } n = j_m \mbox{ and the case-expression may be in a bound or in the in-expression}\\
   & \mbox{and where } y_i  \mbox{ are fresh variables}\\

 \quad & \\
 
\mbox{(ndl)} & (\tchoi~s~t)   \to   s\\
\mbox{(ndr)} & (\tchoi~s~t)  \to   t\\\hline
 
\end{array}\]
\caption{Reduction rules of  \calc} 
\end{figure}


\BD\label{def-red-rules} The reduction rules are defined below in figure \ref{figure-reductions}. 
If the context is important, 
then we denote it as a label of the reduction or state it explicitly. 
Note that for (case), the typical example is written down, where the position of the \tcase~is left open.
 There are two variants, one where the \tcase
is in the in-expression, and one where the \tcase-expression is in the right hand side of a binding. 
An exceptional case, where perhaps a \tletrec-expression has to be omitted, is the case of a constructor 
with zero arguments like $(\tcase~Nil ~\ldots)$.   
\ED 

The union of $ndl, ndr$ is called $(nd)$. 
Reductions are denoted using an arrow with super and/or subscripts: e.g. $\RRAP{llet}$. 
Transitive closure is denoted by a $+$, reflexive transitive closure by a $*$.
 E.g. $\RAP{*}$ is the reflexive, transitive closure of $\to$.  

As a short comment of the reduction rules: 
\BI
\item (lbeta) is a sharing version of beta-reduction
\item (cpn)  is a lazy version of the replacement done by usual beta-reduction, where
 the copy may jump over several indirections. 
\item (case) is the generalized {\tt if} for case analysis of values. To find the value to be analyzed, 
it has to be virtually assembled by following the bindings.  
\item (llet), (lapp), (lcase) are used to adjust the let-environments
\item (nd) is the non-deterministic (erratic) choice.    
\EI

The next definition is intended to formalize the standard reduction. The idea is to find the 
reduction that is outermost, in a reduction context and also necessary for making progress in the 
evaluation. 

\BD
Let $t$ be an  expression. Let  $R$ be the maximal reduction context such that 
$t \equiv R[t']$ for some $t'$.   
The {\em standard redex} and the corresponding standard reduction 
$\RRAP{st}$ is defined by one of the following cases: 

\BI
  \item $t'$ is a choice-expression:  then use (ndr) or (ndl). 
  \item $R = \tletrx{x_1 = t_1,\ldots x_n = t_n}{[]}$, and $t'$ is a {\tt letrec}-expression. Then
  apply (llet) to $R[t']$. 
  \item $R = R_0[(\cdot~t'')]$ where $R_0$ is a reduction context. 
   If $t'$ is a $\tletrec$-expression, then use 
  (lapp) in context $R_0$; If $t'$ is an abstraction, then use 
  (lbeta) in context $R_0$.
   
  \item $R = R_0[\tcase_A~\cdot~alts]$. \\
  If  $t'$ is a $\tletrec$-expression, 
  then use (lcase) in context $R_0$; \\
  If $t'$ is a saturated constructor application, then use (case) in context $R_0$, if it is applicable. 
  
  \item $R = \tletrx{x_1  = [], x_2 = x_1, \ldots, x_j = x_{j-1}, \ldots}{R_1^-[x_j]}]$ where $R_1^-$ is 
  a weak reduction context.\\
  If $t'$ is an abstraction, then 
   use (cpn) as follows: 
  $\tletrx{x_1  = t', x_2 = x_1, \ldots, x_j = x_{j-1}, \ldots}{R_1^-[x_j]}]$ $\to$ 
  $\tletrx{x_1  = t', x_2 = x_1, \ldots, x_j = x_{j-1}, \ldots}{R_1^-[t']}]$. \\
  If $t'$ is a $\tletrec$-expression, then use (llet),(lcase), or (lapp) to flatten the $\tletrec$-expression 
  $t'$ into its superexpression. \\
  If $t'$ is a constructor application, and (case) is applicable to a case-expression in a reduction context, 
  then apply this (case)-reduction. 
  
  \item $R = \tletrx{x_1  = [],x_2 = x_1,\ldots, x_j = x_{j-1},  x_{j+1} = R_{j+1}^-[x_{j}],\ldots}
  {R_{\infty}^-[x_k]}]$ 
  where  $R_i^-, R_{\infty}^-$  are  weak reduction contexts.\\
  If $t'$ is an abstraction, then use (cpn) such that the result is: 
  $R = \tletrx{x_1  = t',x_2 = x_1,\ldots, x_j = x_{j-1},  x_{j+1} = R_{j+1}^-[t'],\ldots}
  {R_{\infty}^-[x_k]}]$.\\
  If $t'$ is a $\tletrec$-expression, then use (llet),(lcase), or (lapp) to flatten the $\tletrec$-expression 
  $t'$ into its superexpression. \\
  If $t'$ is a constructor application, and (case) is applicable to a case-expression in a reduction context, 
  then apply this (case)-reduction.  
\EI
\ED

\BL
For every term $t$: if $t$ has a standard redex, then this redex is unique.
If the standard reduction is not an (nd), then the standard reduction is 
also unique. 
 \EL

\BD  
A standard 
reduction $s \RRAP{st} s_1 \RRAP{st} s_2 \ldots  s_n \RRAP{st} t$
{\em has nd-count} $D$, iff $D$ is the number of (nd)-reductions in it. 
\ED

Note that we use the notion standard reduction also for  non-maximal reductions. 
 
\BD
For a term $t$ and an nd-count $D$, $t\CON_D$ holds if there is some standard-reduction starting with $t$, 
and the reduction has nd-count $D$. 
\ED

Note that a standard reduction for an nd-count $D$ is in general not unique. Note also that there may be 
expressions without a standard redex.

\BD \quad (contextual preorder and equivalence) Let $s,t$ be terms. 
We define: 

\[\begin{array}{lcl} 
  s \le_c t  &  \mbox{ iff }&  \forall C[]. \forall D :  
C[s]\CON_D \impl (\exists B. D \le B \wedge C[t]\CON_B)\\
 s \sim_c t &\mbox{ iff }&  s \le_c t \wedge t \le_c s 
\end{array}\]
\ED

Note that we permit contexts such that $C[s]$ is an open term.

\begin{proposition}
$\le_c$ is a preordering and $\sim_c$ is an equivalence relation.

$s \le_c t$ implies that $C[s] \le_c C[t]$ for all contexts $C[.]$. I.e., $\le_c$ is
a {\em precongruence} on the set of expressions.

$s \sim_c t$ implies that $C[s] \sim_c C[t]$ for all contexts $C[.]$. I.e., $\sim_c$ is
a {\em congruence} on the set of expressions.
\end{proposition}

Note that there are terms $t$ without a standard redex, i.e. the standard reduction stops. 
The reasons could be classified as i) type-error like 
 $(\tcase_A (\lambda x.x) \ldots)$, ii) a kind of non-termination like $\tletrx{x = x}{x}$, 
 iii) as a value or a kind of normal form like \verb|(Cons True Nil)| or $\lambda x. x$.  

The following lemma shows that it is sufficient to use reductions contexts for checking contextual
approximation. 

\begin{lemma} \quad (Context Lemma) Let $s,t$ be terms.
If for all reduction contexts $R$ and all nd-counts $D$: 
$R[s]\CON_D \impl (\exists B. D \le B \wedge R[t]\CON_B)$, then $s \le_c t$.
\end{lemma}

\proof  
We prove the more general claim:  
\BQ
if for all $i$: $s_i, t_i$ satisfy the conditions of the lemma for reduction contexts, then 
 for all multicontexts $C[\cdot_1,\ldots,\cdot_m]$:  
 $C[s_1,\ldots,s_n]\CON_D \impl (\exists B. D \le B \wedge C[t_1,\ldots,t_n]\CON_B$.  
\EQ

Assume this is false. Then there is a multicontext $C$, an nd-count $D$, such that 
$C[s_1,\ldots,s_n]\CON_D$,
and for all $B$ with $D \le B$: $C[t_1,\ldots,t_n]\NOTCON_B$. 
 
We select a multicontext, $C$, terms $s_i, t_i$, and an nd-count $D$, and a corresponding reduction, such that
 the counterexample  is minimal w.r.t. the following lexicographic ordering:
i) the number of reduction steps of $C[s_1,\ldots, s_n]$, ii) the number of holes of $C[\ldots]$.

The search for a standard redex is performed top-down. There are two cases: \\
\BE
\item The search for the reduction context inspects the term in a hole. 
Then we can assume wlog that the first hole is inspected first.   
Hence  $C[\cdot,t_2,\ldots,t_n]$ is a reduction context.  
 Let $C' := C[s_1,\cdot_2,\ldots,\cdot_n]$. Since 
 $C'[s_2,\ldots,s_n] \equiv C[s_1,\ldots,s_n]$, we can select the the same standard
  reduction for nd-count $D$. Since the  number of holes of $C'$ is smaller than the number of holes in $C$,
  we obtain some $B \ge D$ with  $C'[t_2,\ldots,t_n]\CON_B$,
which means $C[s_1,t_2,\ldots,t_n]\CON_B$. Since $C[\cdot,t_2,\ldots,t_n]$ is a reduction
context, the preconditions of the lemma imply that there is some $B' \ge B$ with 
 $C[t_1,t_2,\ldots,t_n]\CON_{B'}$, a contradiction. 

\item The search for the reduction context does not inspect any hole of $C$.
Then $C[s_1,\ldots,s_n]$ as well as $C[t_1,\ldots,t_n]$ can be reduced using the same standard
reduction, since the search for a standard redex takes place only in 
the outer context $C[\ldots ]$.  There are two cases for a reduction: 

If the reduction $C[s_1,\ldots,s_n] \RRAP{st} s'$ is not  (nd), then  this may  
result in $C'[...]$ with more holes, and the holes are filled 
with copies of $s_i,t_i$. Then we get a smaller counterexample since the number of reductions steps is smaller, 
and since non-(nd) standard reductions are unique.

If the reduction $C[s_1,\ldots,s_n] \RRAP{st} s'$ is an (ndl) (or (ndr), respectively), then the reduction of $s'$
 has nd-count $D' = D-1$.  We make the corresponding reduction:  $C[t_1,\ldots,t_n] \RRAP{st,ndl} t'$.
 This is a smaller counterexample; hence we get a 
contradiction also in this case. 
\EE

\BD A {\em program transformation} is a relation $T$ between programs (expressions). 
A program transformation $T$ is called {\em correct}, iff for all expressions $P,P'$: 
$P~T P'$ implies $P \sim_c P'$.  
\ED

The reductions rules in definition \ref{def-red-rules} define corresponding 
program transformations if they are allowed
in arbitrary contexts.

\BD
Let an {\em internal reduction} be a non-standard reduction 
that takes place  within a 
reduction context. Usually, this is denoted by the label $i$ on the reduction arrow.
\ED 

We define complete sets of commuting and forking diagrams adapted from \cite{kutzner:99,kutzner:00}.
In the following definition we use a notation for rewrite rules on reduction sequences.
For example   $\RAP{(i,llet)} \cdot \RAP{(st,a)} ~\leadsto~  \RAP{(st,a)} \cdot \RAP{(i,llet)}$, where
$a$ is a reduction type. 
 The $\cdot$ on the left hand side
is like a joker, and the $\cdot$ on the right hand side can be seen as an existentially quantified 
term.  

\BD  Assume given a reduction type (red) and a set of (complementary) reduction types $T$, where  
 the base calculus reduction types are contained  in $T$, as well as (red).  

A {\em complete set of commuting diagrams} for a reduction (red) is a set of rewrite rules on 
reduction sequences of the
form $$\RRAP{i,red} . \RRAP{st,a_1} \ldots \RRAP{st,a_k}  ~~Ê \leadstoÊ~~
\RRAP{st,b_1} \ldots \RRAP{st,b_m} .
\RRAP{i,c_1} . \ldots . \RRAP{i,c_h},$$  
where $c_i \in T$, such that for 
 every reduction sequence $s \RRAP{i,red} . \RRAP{st,*} t$: 
 Either it can be transformed using one of the meta-reductions into another
 reduction sequence from $s$ to $t$,  such that at least  $\RRAP{i,red}$ can be replaced.
 Or $. \RRAP{st,*} t$ can be prolonged into a longer standard reduction sequence
 $\ldots \RRAP{st,*} t \RRAP{st,+} t'$, such that it can be replaced as above.
\vspace{0,3cm}

A {\em complete set of forking diagrams} for a reduction (red) is a set of rewrite rules 
on reduction sequences of the
form $$\LLAP{st,a_1} \ldots \LLAP{st,a_k} . \RRAP{i,red} ~~Ê\leadstoÊ~~Ê 
\RRAP{i,c_1} . \ldots . \RRAP{i,c_h} .
\LLAP{st,b_1} \ldots \LLAP{st,b_m},$$  
where $c_i \in T$, such that:
 Either
 every reduction sequence $s \LLAP{st,*} . \RRAP{i,red} t$ can be transformed into another
 reduction sequence between $s$ and $t$, such that at least  $\RRAP{i,red}$ is replaced. 
 Or $s \LLAP{st,*} . $ can be prolonged into a reduction sequence
 $s' \LLAP{st,+} s \LLAP{st,*} .~$, such that it can be replaced as above.
 
We also use reductions not in the base calculus as internal reductions. 
 \ED

It is intended that the corresponding meta-rewriting on reduction sequences terminates, which has to be proved for 
every such complete set. The complete sets of commuting (forking) diagrams are not unique.   

Note that in many cases, the forking diagrams can be derived from the commuting diagrams.

\BL\label{lemma-no-internals}
For every reduction that is not a (llet) or (cp)-reduction, i.e.,  
 $a \in  \{(nd),(lbeta),(lapp),(lcase),(case)\}$, there are no 
internal reductions. 
This means, every internal $a$-reduction with 
$a \in  \{(nd),(lbeta),(lapp),(lcase),(case)\}$ 
is a standard reduction.
\EL  
\proof
By inspecting all the  finitely many cases.

\begin{proposition}
If $s \RRAP{a} t$, where $a \in \{(lbeta),(lapp),(lcase),(case)\}$, then 
$s \sim_c t$. 

I.e., all the program transformations defined by one of the reductions \{(lbeta),(lapp),(lcase),(case)\} 
are  correct. 
\end{proposition}
\proof
Let $s' \RRAP{a,[]} t'$ by a (a)-reduction on the surface with 
$a \in \{(lbeta),(lapp),(lcase),(case)\}$.

We show $s' \le_c t'$ exploiting the context lemma. 
Let $R$ be a reduction context. Then $R[s'] \RAP{i,a} R[t']$ is not possible by 
Lemma \ref{lemma-no-internals}.
Then $R[s'] \RAP{st,a} R[t']$ by a unique standard
 reduction, hence if there a reduction for $R[s']$ with nd-count $D$, there is also 
 one for $R'[t']$ with nd-count $D$. 
 The context lemma now shows that $s' \le_c t'$.  

To show $t' \le_c s'$ using the context lemma is similar: 
If there a standard reduction for 
$R[t']$ with nd-count $D$, there is also  one for $R[s']$ with nd-count $D$. 
 The context lemma now shows that $t' \le_c s'$. 
  
Hence we have shown $s' \sim_c t'$. Since $\sim_c$ is a congruence, 
we have also that  $C[s'] \sim_c C[t']$ for an arbitrary context $C$. Hence the proposition holds.

\section{Correctness of the reduction (llet)}

The union of the reductions (llet),(lapp),(lcase) is denoted as (lll). The reduction $lll^+$  means 
a reduction sequence consisting only of  (lll)-reductions of length at least 1. Accordingly $lll^*$ is defined
as any number of (lll)-reductions.  $(i,llet)^{0\vee 1}$ means
no reduction or 1 reduction $(i,llet)$. In the following two lemmas, $a$ stands for an arbitrary reduction
 $\RRAP{a}$. 

\begin{lemma}
A complete set of commuting diagrams  for (llet) is: 
\begin{itemize}
\item $\RRAP{(i,llet)} \cdot \RRAP{(st,a)} ~\leadsto~  \RRAP{(st,a)} \cdot \RRAP{(i,llet)}$
\item $\RRAP{(i,llet)} \cdot \RRAP{(st,a)} ~\leadsto~  \RRAP{(st,a)} \cdot \RRAP{(st,llet)}$
\item  $\RRAP{(i,llet)} \cdot \RRAP{(st,lll^+)}$   
$~\leadsto~$
 $\RRAP{(st,lll^+)}  \cdot  \RRRAP{(i,llet)^{0\vee 1}}$ 
\end{itemize}
\end{lemma}

\begin{lemma}
A complete set of forking diagrams  for (llet) is: 
\begin{itemize}
\item $\LLAP{(st,a)} \cdot \RRAP{(i,llet)} ~\leadsto~  \RRAP{(i,llet)} \cdot \LLAP{(st,a)}$
\item $\LLAP{(st,llet)} \cdot \LLAP{(st,a)} \cdot \RRAP{(i,llet)} ~\leadsto~   \LLAP{(st,a)}$
\item  $\LLLAP{(st,lll^+)} \cdot  \RAP{(i,llet)}$   
$~\leadsto~$
 $\RAP{(i,llet} \cdot \LLLAP{(st,lll^+)}$ 
\item  $\LLLAP{(st,lll^+)} \cdot  \RAP{(i,llet)}$   
$~\leadsto~$
 $\LLLAP{(st,lll^+)}$ 
\end{itemize}
\end{lemma}


\begin{proposition}
If $s \RRAP{(i,llet)} t$, then $s \sim_c t$.\\ I.e. (llet) is a correct program transformation 
in any context.
\end{proposition}
\proof
First we assume that the reduction is on top level. 

To use the context lemma, we have to show what happens in a  reduction  context. I.e. assume that
$s \equiv R[s']$ and $s'$ is the llet-redex.     

Using the forking diagrams, it is possible to construct from a standard reduction 
of $s$ a standard reduction of $t$ with the same nd-count.  The context lemma then shows
that $s \le_c t$. 

Using the commuting diagrams, it is possible to construct from a standard reduction 
of $t$ a standard reduction of $s$ with the same nd-count by shifting the $\RRAP{(i,llet)}$ to the right. 
The context lemma then shows
that $t \le_c s$.

Together, this means $s \sim_c t$. 

Finally, the congruence property of $\sim_c$ implies that a (llet) can be applied everywhere in a term.  
\QED

%
%
%

\section{Garbage Collection: ldel}

Garbage collection in the calculus has two forms, a non-cyclic one,
and the other one that also collects cyclic references: 

The noncyclic reduction (ldel)  is defined as : 

\[\begin{array}{l@{\quad}l}
 (ldel)&    \tletrx{x = s}{t} \to t  \mbox{ if } x  \mbox{ does not occur in }  t \\
 (ldel)&    \tletrx{x = s,E}{t} \to \tletrx{E}{t}  \mbox{ if } x  \mbox{ does not occur in }  t,E \\
\end{array}\]

The  cyclic reduction (ldelcyc) consisting of (ldelcyc1), (ldelcyc2) is defined as : 
    
\[
\begin{array}{ll}
\mbox{(ldelcyc1)}~ &  
   \tletrx{x_1 = s_1, \ldots, x_m = s_m}{t} \to \tletrx{x_j = s_j, \ldots, x_m = s_m}{t} \\
  
& \mbox{ if } x_i \mbox{ for } 1 \le i \le j-1 \mbox{ does not occur in } s_j,\ldots,s_m, t 
\mbox{ and } m > 1\\
\mbox{(ldelcyc2)}~ &  
      \tletrx{x_1 = s_1, \ldots, x_m = s_m}{t} \to t \\
& \mbox{ if } x_i \mbox{ for } 1 \le i \le m \mbox{ does not occur in } t 
 
\end{array}\]

Here we show the correctness of (ldel). 

\begin{lemma}\label{lemma-ldel-commuting}
A complete set of commuting diagrams for (ldel) is: 
\begin{itemize}
\item $\RRAP{(ldel)} \cdot \RRAP{(st,a)} ~\leadsto~ \RRAP{(st,a)} \cdot \RRAP{(ldel)}$   
\item $\RRAP{(ldel)}  ~\leadsto~ \RRAP{(st,lll^+)} \cdot \RRAP{ldel}$
\item $\RRAP{(ldel)} \cdot \RRAP{(st,lll^+)}  ~\leadsto~ \RRAP{(st,lll^*)} \cdot \RRAP{(ldel)}$
 
\end{itemize}
\end{lemma}

As an example for computing commuting diagrams, we show one case:
We write \verb|\| instead of $\lambda$. 
\BEX
We compute the overlap of an (ldel)-redex and a standard (lapp)-redex. If the overlap is trivial, then it is
not hard to see that the reductions commute, including the property ``standard``.

In the case of a proper overlap, the redex and the corresponding reduction is as follows:  \\
\verb|((letrec x = c in \y.y) d)| $\RRAP{ldel}$ \verb| (\y.y  d)| \\
On the other hand, if first the (lapp) rule is applied, then:\\
\verb|((letrec x = c in \y.y) d)| $\RRAP{st,lapp}$ \verb|(letrec x = c in (\y.y  d))| \\
  $\RRAP{ldel}$ \verb| (\y.y  d)| \\
This is covered by the rule
 $\RRAP{(ldel)}  ~\leadsto~ \RRAP{(st,lll^+)}  \cdot \RRAP{ldel}$.
\EEX
 
\begin{lemma}
A complete set of forking diagrams for (ldel) is: 
\begin{itemize}
\item $\LLAP{(st,a)}\cdot \RAP{(ldel)} ~\leadsto~  \RAP{(ldel)} \cdot \LLAP{(st,a)}$  
\item $\LLLAP{(st,lll^+)} \cdot \RAP{(ldel)}
  ~\leadsto~ \RAP{(ldel)}$ 
\item $\LLLAP{(st,lll^+)}\cdot \RAP{(ldel)}
  ~\leadsto~ \RAP{(ldel)} \cdot  \LLAP{(st,lll^*)}$
\end{itemize}
\end{lemma}

\begin{lemma}
There are no infinite $lll$-reductions
\end{lemma}
\proof
This can be shown by a natural-number valuation of expressions similar as in \cite{kutzner:00}, 
which is strictly decreasing in every reduction step.  
\QED

\begin{proposition}
If $s \RRAP{(ldel)} t$, then $s \sim_c t$.
\end{proposition}
\proof (sketch)\\
Follows by induction on the length of reductions from the context lemma, and since 
there are no infinite (lll)-reduction sequences. 
\QED

%
%
%

\section{Copying variables}

This section contains the reduction (lcv) which is like compressing references used in \tletrec s. It can also
be described as removing indirections. 
 
\[\begin{array}{ll}
\mbox{(lcv)} &  \tletrx{x = y,E}{C[x]} \to \tletrx{x = y, E}{C[y]} \\
 \mbox{(lcv)} &   \tletrx{x_1 = y,x_2 = C[x_1], E}{t} \\
 &  \to  \tletrx{x_1 = y, x_2 = C[y], E}{t}
 \end{array}\]

\begin{lemma}
A complete set of commuting diagrams for (lcv) is: 

\begin{itemize}
\item $\RAP{(lcv)} \cdot \RAP{(st,a)} ~\leadsto~ \RAP{(st,a)} \cdot \RAP{(lcv)}$
\item $\RAP{(lcv)} \cdot \RAP{(st,cpn)} ~\leadsto~ \RAP{(st,cpn)} \cdot \RAP{(lcv)} \cdot \RAP{(lcv)}$.
\item $\RAP{(lcv)} \cdot \RAP{(st,a)} ~\leadsto~ \RAP{(st,a)}$, where $a \in \{case,cpn,ndr,ndl\}$.  

\end{itemize}
\end{lemma}

\begin{lemma}
A complete set of forking diagrams for (lcv) is: 
\begin{itemize}
\item $\LLAP{(st,a)} \cdot \RRAP{(lcv)} ~\leadsto~  \RRAP{(lcv)}  \cdot \LLAP{(st,a)}$
\item $\LLAP{(st,cpn)} \cdot \RRAP{(lcv)} ~\leadsto~  \RRAP{(lcv)}  \cdot  \RRAP{(lcv)} 
 \cdot \LLAP{(st,cpn)}$
\item  $\LLAP{(st,a)} \cdot \RRAP{(lcv)} ~\leadsto~  \LLAP{(st,a)}$ for $a \in \{cp,case,ndl,ndr\}$.

\end{itemize}
\end{lemma}

\begin{proposition}
If $s \RRAP{(lcv)} t$, then $s \sim_c t$. 

I.e., (lcv) is a correct program transformation in any context.
\end{proposition}

The proof uses the context lemma, and the complete set of commuting and forking diagrams to 
meta-reduce reduction sequences. 

%
%
%

\section{Contextual equivalence of copy reductions}

The required diagrams and the proof of correctness of non-standard copy reductions 
are complex. Only the  complete set of commuting diagrams are presented.  

For this rule we require a special class of contexts: surface contexts: 
Surface contexts define expressions with holes not in the body of an abstraction.  
  
\BD
\begin{eqnarray*}
 S &::= &~[]~|~(S~E)~|~(E~S)~|~(\tcase_A~S~alts)~|~(\tcase_A~E~\ldots  (p \to S)  \ldots ) \\ 
    &  &~|~(\tchoi~E~S)~|~(\tchoi~S~E)~\\ 
    &  &~|~(\tletrx{\ldots}{S}~|~ (\tletrx{\ldots, x_i = S, \ldots}{E}   
\end{eqnarray*}
where $E$ stands for an expression. $S$ is called {\em surface context}. 
\ED

We consider the following  atomic copy reductions: 

\[\begin{array}{ll}
\mbox{(cp)} & \tletrx{x_1 = s_1,\ldots , x_n = s_n}{C[x_1]}  \\
& \quad \to \tletrx{x_1 = s_1,\ldots , x_n = s_n}{C[s_1]} \\
& \mbox{where } s_1 
     \mbox{ is an abstraction}\\
\mbox{(cp)} & \tletrx{x_1 = s_1,\ldots , x_n = s_n}{s}  \\
& \quad \to  \tletrx{x_1 = s_1,\ldots , x_j = C[s_1],\ldots, x_n = s_n}{s} \\
& \mbox{where } s_1 
     \mbox{ is an abstraction} \\
     & \mbox{and where } s_j \equiv C[x_1]\\
\end{array}\]

We distinguish the (cp)-reduction into two subreductions: 
If the target occurrence of the variable is in a surface context, then (cpt), otherwise it is a (cpd). 
Equivalently, it is a (cpd) iff the target variable is within an abstraction. 

Thus $(\tletrx{x = s, E}{D[\lambda z. C[x]]}) \to  (\tletrx{x = s, E}{D[\lambda z. C[s]]})$ 
is a reduction of type (cpd).

\begin{lemma}
A complete set of commuting diagrams for (cpt), (cpd) is: 

\begin{itemize}
\item $\RAP{(i,cpt)} \cdot \RAP{(st,a)} ~\leadsto~ \RAP{(st,a)} \cdot \RRAP{(\{i,st\},cpt)}$
\item $\RAP{(i,cpt)} \cdot \RAP{(st,a)} ~\leadsto~ \RAP{(st,a)}$, where $a \in\{case, ndr,ndl\}$. 

\item $\RAP{(i,cpd)} \cdot \RAP{(st,a)} ~\leadsto~ \RAP{(st,a)} \cdot \RAP{(i,cpd)}$
\item $\RAP{(i,cpd)} \cdot \RAP{(st,cpn)} ~\leadsto~ \RAP{(st,cpn)} \cdot \RAP{(i,cpd)} \cdot \RAP{(i,cpd)}$
\item $\RAP{(i,cpd)} \cdot \RAP{(st,a)} ~\leadsto~ \RAP{(st,a)}$, where $a \in\{case, ndr,ndl\}$. 
\item $\RAP{(i,cpd)} \cdot \RAP{(st,lbeta)} ~\leadsto~ \RAP{(st,lbeta)} \cdot \RRAP{(\{i,st\},cpt)}$.
\end{itemize}
 
\end{lemma}

This is sufficient to show that the (cp)-reductions retain contextual equivalence by a meta-reduction 
on reduction sequences.  

\begin{proposition}
If $s \RRAP{(cp)} t$, then $s \sim_c t$. 

I.e., (cp) is a correct program transformation in any context.
\end{proposition}

In summary, we can prove: 

\begin{theorem}
All the reductions of the base calculus with the exception of (ndr), (ndl) are correct program transformations 
\end{theorem}

It is obvious that (ndr), (ndl)  are not correct as program transformations, since $(\tchoi~True~False)$ may reduce
to $True$, but $True$ is not equivalent to $(\tchoi~True~False)$.

\section{Unique Copy: Inlining}

If a letrec-bound variable occurs only once, then it is possible to replace this variable by the bound 
expression and to remove the binding: 
 
\begin{tabular}{ll}
 (ucp)~  &  $\tletrx{x = s, E}{C[x]} \to  \tletrx{E}{C[s]}$, where $C[]$ is a surface \\
    &   context, $s$ arbitrary, $x$ has exactly one occurrence in $C[x]$ and no \\
  &  occurrence in $E$ nor in $s$.\\ 
  (ucp)~  &  $\tletrx{x = s}{C[x]} \to  C[s]$, where $C[]$ is a surface context, $s$ arbitrary,\\
  &    and $x$ has exactly one occurrence in $C[x]$ and no occurrence in $s$.\\ 
 (ucp)~Ê &   $\tletrx{x = s, y = C[x], E}{t} \to  \tletrx{y = C[s], E}{t}$, where $C[]$  \\
               &  is a surface context, $s$ arbitrary, $x$ has exactly one occurrence in $C[x]$  \\
               & and no occurrence in $E$,$s$ and $t$. 
\end{tabular}

Note that if $s$ is an abstraction, then the rule is a combination
of (cp) and (ldel).

\begin{lemma}
A complete set of commuting diagrams for (ucp) is: 
\begin{itemize}
\item $\RAP{(ucp)} \cdot \RAP{(st,a)} ~\leadsto~ \RAP{(st,a)} \cdot \RAP{(ucp)}$
\item $\RAP{(ucp)} \cdot \RAP{(st,a)} ~\leadsto~ \RAP{(st,a)} \cdot \RAP{(ldel)}$ for $a \in \{case,ndr,ndl,cpn\}$
\item $\RAP{(ucp)} \cdot \RAP{(st,a)} ~\leadsto~ \RAP{(st,a)}$ for $a \in \{case,ndr,ndl\}$
\item $\RAP{(ucp)} \cdot \RAP{(st,lll^*)}~\leadsto~ \RAP{(st,lll^*)} \cdot \RAP{(ucp)}$ \\
where the extreme~
case $\RAP{(ucp)} \leadsto~  \RAP{(ucp)}$ is excluded.
\item $\RAP{(ucp)} ~\leadsto~ \RAP{(st,cpn)} \cdot \RAP{(ldel)}$
\end{itemize}
\end{lemma}

\begin{proposition}
If $s \RRAP{(ucp)} t$, then $s \sim_c t$. 

I.e., (ucp) is a correct program transformation in any context.
\end{proposition}

\section{Conclusion}

The rewriting based method of computing complete sets of commuting (resp. forking) diagrams to prove 
program transformations 
to be correct   is demonstrated 
to be successful. We are able to  show that all deterministic  reduction rules in 
the rather complex lambda calculus \calc and also some other rules are correct.  
A general automatic method to compute diagrams by checking all non-trivial overlaps would be a valuable tool
and deserves further research efforts.

\newcommand{\etalchar}[1]{$^{#1}$}

\end{document}